# On the calculation of diffusion coefficients in confined fluids and interfaces with an application to the liquid-vapor interface of water

Pu Liu, Edward Harder, and B. J. Berne

*Department of Chemistry and Center for Bimolecular Simulation*

*Columbia University*

*3000 Broadway, New York, New York 10027*

## Abstract

We propose a general methodology for calculating the self-diffusion tensor from molecular dynamics for a liquid with a liquid-gas or liquid-solid interface. The standard method used in bulk fluids, based on computing the mean square displacement as a function of time and extracting the asymptotic linear time dependence from this, is not valid for systems with interfaces or for confined fluids. The method proposed here is based on imposing virtual boundary conditions on the molecular system and computing survival probabilities and specified time correlation functions in different layers of the fluid up to and including the interfacial layer. By running dual simulations, one based on MD and the other based on Langevin dynamics, using the same boundary conditions, one can fit the Langevin survival probability at long times to the MD computed survival probability, thereby determining the diffusion coefficient as a function of distance of the layers from the interface. We compute the elements of the diffusion tensor of water as a function of distance from the liquid vapor interface of water. Far from the interface the diffusion tensor is found to be isotropic, as expected, and the diffusion coefficient has the value $D \approx .22 \text{Å}^2/\text{psec}$ in agreement with what is found in the bulk liquid. In the interfacial region the diffusion tensor is axially anisotropic, with values of $D_\parallel \approx .8$Å$^2$/psec and $D_\perp \approx .5$Å$^2$/psec for the components parallel and normal the interface surface respectively. We also show that diffusion in confined geometries can be calculated by imposing appropriate boundary conditions on the molecular system and computing time correlation functions of the eigenfunctions of the diffusion operator corresponding to the same boundary conditions.

# I. INTRODUCTION

Molecular dynamics (MD) computer simulations provide a powerful tool for analyzing macroscopic transport coefficients like diffusion at a microscopic level. The use of computer simulation to analyze the properties of water is a common tool[1,2] and the calculation of the self-diffusion coefficient from simulation has yielded consistent results to experiment for modern water models.[3] In uniform fluids, two methods have been used to compute the self-diffusion coefficient from molecular dynamics trajectories. In one of these methods, one computes the mean-square displacement (MSD) of the molecules, $\langle \Delta r(t)^2 \rangle$ as a function of time and fits this to the asymptotic time dependence predicted from the diffusion equation,

$$\langle \Delta r(t)^2 \rangle \to 2dDt. \tag{1}$$

This is the Einstein relation, from which the diffusion coefficient can in turn be expressed in terms of the MD calculated autocorrelation function of the velocity, giving the Green-Kubo relation,

$$D = \frac{1}{d} \int_0^\infty dt \langle \mathbf{v}(t) \cdot \mathbf{v}(0) \rangle \tag{2}$$

where $d$ is the dimensionality of the system. These methods are derived from the ordinary diffusion equation with free boundary conditions and are thus only suitable for calculations in homogeneous fluids.

If, instead, we wish to determine the diffusion coefficient of liquids confined to cavities, or in inhomogeneous regions such as, for example, either the air-water interface or the interface between water and various solids,[4–9] neither the Einstein relation nor the Kubo relation are valid approaches. In confined liquids the mean square displacement will be bounded by the size of the confined region and it will be difficult to unambiguously find the diffusion coefficient. In inhomogeneous systems, molecules will stay in the region of interest only for a finite time and then will explore other regions. Since the diffusion coefficient will be different for different regions, the time dependence of the MSD, computed for particles initially in the region of interest, will only become linear at times long enough for the molecules to sample all regions, and then its slope will give the diffusion coefficient averaged over all regions. How then will we be able to determine the diffusion coefficient for the region of interest. Moreover for such systems the spatial distribution at equilibrium will not be uniform and the appropriate phenomenological equation will be the Smoluchowski equation and not the free diffusion equation from which the Einstein relation is derived. In this paper we present various strategies for determining the diffusion coefficients in such systems. In the case of a liquid vapor interface, the asymmetry of the interface region introduces unique features in interfacial self diffusion. Special care is required in the analysis of such properties and necessitates an accurate determination of the diffusion tensor in the interfacial region.

Since the simple diffusion equation is not valid for confined or inhomogeneous systems a new approach is necessary. In the methods proposed here we make use of molecular dynamics with virtual boundary conditions. Lacking new methods, Eqs. (1) and (2) have been used in the past and continue to be applied to the interfacial region of liquids. When used for interfaces, the system is divided into layers. The MSD for particles in any layer is calculated even if their trajectories carry them out of the layer. If the layer is thin, the linear time dependence will reflect diffusion in neighboring layers as well. Moreover in the interfacial layer the molecules will be diffusing according to the Smolochowski equation, not the free diffusion equation, so that use of the Einstein relation to determine the diffusion coefficient is questionable.

This article introduces a new methodology for calculating the diffusion tensor in an inhomogeneous system in Section II and applies this methodology to the study of the air-water interface in Section III. The methodology is different for diffusion in a direction perpendicular to the interface than for a direction parallel to the interface. In both cases we introduce virtual absorbing boundaries at the boundaries of layers. For determination of the diffusion coefficient parallel to the interface, $D_{\parallel}$, we generalize the Einstein relation and show that the MSD of particles that remain in the layer asymptotically varies as $\rightarrow 2P(t)D_{\parallel}t$, so that an MD determination of the survival probability in the layer, $P(t)$, and the MSD allows determination of the parallel component of the diffusion tensor. For determination of the diffusion coefficient perpendicular to the interface we propose a dual simulation technique. The gist of this method is to perform MD with virtual absorbing boundary conditions at each layer and to perform a sequence of Langevin dynamics sim-ulations each with the potential of mean force determined from the MD, but with different values of friction coefficient $\zeta_\perp$ characterizing motion in a direction perpendicular to the interface. One can find the value of the friction coefficient that gives the best fit of the Langevin survival probability in a layer to that found by MD for that layer. $D_\perp$ can then be found from the Stokes-Einstein relation, $D_\perp = k_B T/\zeta_\perp$. This "Dual Simulation Method" requires a rather subtle definition of the boundaries for the Langevin dynamics. We also show that, in some cases, we can determine the diffusion coefficient for fluids in confined geometries as well as for $D_\perp$ in layers with absorbing boundary conditions from a computation of the autocorrelation function of the eigenfunctions of the diffusion operator corresponding to these boundary conditions. The various approaches are tested in a variety of ways.

This study was motivated by our interest in water hydrogen bond dynamics in the liquid-vapor interface.[10] Previous studies have shown that diffusion makes an important contribution to the hydrogen-bond dynamics.[11] Before understanding hydrogen bond dynamics in interfaces we must understand diffusion there. Thus we apply our methodology to the liquid-vapor interface of liquid water.

## II. METHODOLOGY

Since the presence of the interface breaks the symmetry of the liquid, diffusion in such a system is described by an anisotropic Smoluchowski equation:

$$\partial_t p(\mathbf{r}, t | \mathbf{r_0}, t_0) = \nabla \cdot \mathbf{D} \cdot [\nabla + \beta(\nabla W(z))] p(\mathbf{r}, t | \mathbf{r_0}, t_0) \tag{3}$$

where $p(\mathbf{r}, t | \mathbf{r_0}, t_o)$ is the conditional probability distribution function, and $\mathbf{D}$ is the diffusion tensor. Taking the z-axis to be perpendicular and the x and y axes to lie parallel to the interface, a system with a flat interface will be axially symmetric around the z-axis. The equilibrium density profile along the z direction is given by $\rho(z) = \rho_0 \exp(-\beta W(z))$, where $W(z)$ is the potential of mean force (PMF). The density $\rho(z)$ is easily determined by MD and from this the PMF can be determined. The diffusion tensor will be diagonal in this frame with $D_{zz} \neq D_{xx} = D_{yy}$. In general the diffusion coefficients will be a function of $z$; however, if we study molecules in sufficiently thin layers parallel to the interface, we expect the diffusion coefficients not to vary with z within a layer, but of course they may vary from layer to layer, especially as we approach the interfacial region. The Smoluchowski equation will then be separable in each layer. In this case the Smoluchowski equation in each layer will be separable in x, y and z. The original equation (3) can be decomposed into three independent equations:

$$\partial_t p(x,t|x_0,t_0) = D_{xx}\partial_x^2 p(x,t|x_0,t_0), \tag{4}$$

$$\partial_t p(y,t|y_0,t_0) = D_{yy}\partial_y^2 p(y,t|y_0,t_0), \tag{5}$$

$$\partial_t p(z,t|z_0,t_0) = D_{zz}\partial_z \left[\partial_z + \beta(\partial_z W(z))\right] p(z,t|z_0,t_0), \tag{6}$$

where,

$$p(\mathbf{r},t|\mathbf{r_0},t_o) = p(z,t|z_o,t_o)p(y,t|y_o,t_o)p(x,t|x_o,t_o). \tag{7}$$

In order to analyze the dependence of diffusion on position relative to the interface we derive and compute the components of the diffusion tensor for finite regions that range from the bulk water phase to the interface. Any molecule initially present in a specified region that passes the boundaries no longer contributes to the calculation. In the coordinates parallel to the interface ($x$ and $y$) water diffusion is unbounded. In the $x$ direction, we can postulate the boundary condition:

$$p(x \to \infty, t) = 0, \tag{8}$$

and the initial condition:

$$\lim_{t \to t_0} p(x,t|x_0,t_0) = \delta(x - x_0). \tag{9}$$

The same can be done for diffusion along the $y$-direction. Without loss of generality we take $t_0 = 0$ in the following. Combined with Eqs. (4) and (5), we can get the familiar results for diffusion in the x and y directions:

$$p(x,t|x_0,0) = \frac{1}{\sqrt{4\pi D_{xx}t}} \exp\left[-\frac{(x-x_0)^2}{4D_{xx}t}\right], \tag{10}$$

$$p(y,t|y_0,0) = \frac{1}{\sqrt{4\pi D_{yy}t}} \exp\left[-\frac{(y-y_0)^2}{4D_{yy}t}\right]. \tag{11}$$

A very similar approach can be used for fluids in a confined geometry. For example for a uniform fluid in a rectangular box with reflecting boundary conditions, one may find the eigenfunctions and eigenvalues of the diffusion operator subject to reflecting boundary conditions at the walls. Then the decay of the autocorrelation functions are exponential with well defined time constants. MD of a fluid between reflecting walls can then be used to calculate these correlation functions and from their exponential decay at long times one can find the diffusion coefficients. In spherical cavities, one can solve the diffusion equation with reflecting boundary conditions. The eigenfunctions are spherical Bessel functions multiplied by spherical harmonics.

### A. Diffusion coefficients parallel to the interface

In the water-vapor system, we apply virtual absorbing boundary conditions to the MD to localize the contributions to the diffusion coefficient to our specified finite regions. These are virtual in that they are applied only to the analysis of MD trajectories and not to the generation of these trajectories. The eigenfunctions $\Psi_n$, of Eq. (6) subject to these boundary conditions and the evaluation of the perpendicular component of the diffusion tensor, $D_\perp = D_{zz}$, that follows will be discussed shortly. First we will focus on the simpler problem of evaluating the parallel component, $D_\parallel$, of the diffusion tensor, which is equal to either $D_{xx}$ or $D_{yy}$.

The probability for a particle to remain in this region, $a \leq z \leq b$, can be determined by integrating the conditional probability over $x$, $x_0$, $y$, $y_0$, $z_0$ and $z$:

$$P(t) = \int_a^b dz \int_a^b dz_0 \, p(z,t|z_0,0) \, p(z_0). \tag{12}$$

In order to calculate the components $D_{xx}$ and $D_{yy}$ we formulate the mean square displacement for a finite region. In the $x$ direction the mean square displacement, $\langle \Delta x(t)^2 \rangle_{\{a,b\}}$, of particles that remain in region $\{a,b\}$ is:

$$\begin{aligned}\langle \Delta x(t)^2 \rangle_{\{a,b\}} &= \int d\mathbf{r} \int d\mathbf{r_0} p(\mathbf{r},t|\mathbf{r_0},t_o) p(\mathbf{r_o})(x-x_o)^2 \\ &= 2P(t)D_{xx}t.\end{aligned} \tag{13}$$

It is important to note that $\langle \Delta z^2(t) \rangle_{a,b} \neq 2 D_{zz} P(t) t$.

The mean square displacement and the survival probability $P(t)$ are calculated from the simulation as follows. Let $S(t, t+\tau)$ designate the set of all particles that stay in the layer $\{a, b\}$ during the time interval between $t$ and $t + \tau$, let $N(t, t + \tau)$ be the number of such particles, and finally let $N(t)$ designate the number of particles in the layer at time $t$. Then we compute:

$$\langle \Delta x(\tau)^2 \rangle_{\{a,b\}} = \frac{1}{T} \sum_{t=1}^{T} \frac{1}{N(t)} \sum_{i \epsilon S(t,t+\tau)} (x_i(t+\tau) - x_i(t))^2, \tag{14}$$

and

$$P(\tau) = \frac{1}{T} \sum_{t=1}^{T} \frac{N(t, t+\tau)}{N(t)}. \tag{15}$$

In the foregoing $T$ is the total number of time steps averaged over. Thus we can obtain the expression for $D_{xx}$ and similarly for $D_{yy}$:

$$D_{xx}(\{a,b\}) = \lim_{\tau \to \infty} \frac{\langle \Delta x(t)^2 \rangle_{\{a,b\}}}{2\tau P(\tau)}. \tag{16}$$

It is important to note that the correlation functions defined in Eq. (14) and Eq. (15) are not ordinary two point correlation functions because the requirement that the particle be in the layer for the whole time between $t$ and $t + \tau$ makes this a function of the whole history in this time interval and not just on its values at $t$ and $t + \tau$.

### B. The diffusion coefficient perpendicular to the interface

It remains to discuss methods used to evaluate the component of the diffusion tensor perpendicular to the interface, $D_\perp = D_{zz}$. For finite regions sufficiently far away from the interface, the density is uniform, the potential of mean force, $W$, is zero, and the Smoluchowski equation, Eq. (6), reduces to the diffusion equation,

$$\partial_t p(z, t | z_0, t_0) = D_{zz} \partial_z^2 p(z, t | z_0, t_0). \tag{17}$$

We can derive an analytical expression for the diffusion coefficient with absorbing boundary conditions by applying the commonly used Smoluchowski boundary condition defined by $p(z = a, t) = 0$ and $p(z = b, t) = 0$. The initial condition is $p(z, t_0 | z_0, 0) = \delta(z - z_0)$. Using the standard method of separation of variables,[12] we set $p(z,t|z_0,0) = f(z)g(t)$ and solve Eq. (17) to obtain the solution of the form:

$$p(z,t|z_0,0) = \sum_{n=1}^{\infty} \Psi_n^*(z)\Psi_n(z_0)\exp(-(n\pi/L)^2 D_{zz}t), \quad (18)$$

where,

$$\Psi_n(z) = \sqrt{2/L}\sin(n\pi(z-a)/L) \quad \text{n=1,2...}, \quad (19)$$

are eigenfunctions of the diffusion operator satisfying the boundary conditions, where $L = b - a$.

The autocorrelation of any such eigenfunction is easily determined to be,

$$\langle \Psi_n(z(t))\Psi_n^*(z(0))\rangle = \int_a^b dz \int_a^b dz_0\, p(z,t|z_0,t_0)\,p(z_0)\Psi_n(z)\Psi_n^*(z_0)$$
$$= \frac{1}{L}e^{-(n\pi/L)^2 D_{zz}t}. \quad (20)$$

Here $p(z_0)$ is the initial normalized probability distribution along $z$ in the interval, which is a uniform distribution ($=1/L$) in the bulk water regions.

The autocorrelation of the eigenfunction is easily determined from the simulation as,

$$\langle \Psi_n(z(t))\Psi_n^*(z(0))\rangle_{\{a,b\}} = \frac{1}{T}\sum_{t=1}^{T} \frac{1}{N(t)} \sum_{i\epsilon S(t,t+\tau)} \Psi_n(z_i(t+\tau))\Psi_n^*(z_i(t)), \quad (21)$$

where once again $S(t,t+\tau)$ designates the set of all particles that stay in the layer $\{a,b\}$ during the time interval between $t$ and $t+\tau$, $N(t)$ is the number of particles in the layer at time $t$, and $T$ is the total number of time steps averaged over.

Near the interface one must use the Smulochowski equation. Although not applied in this paper, the eigenfunctions of for this equation (after being made self-adjoint) can easily be found numerically. Evaluation of the autocorrelation function will again decay exponentially, allowing one to determine the diffusion coefficient as above.

Another route to determining the diffusion coefficient, and the one adopted here, avoids solving Eq. (6) directly, involves running Langevin dynamics simulations for the motion perpendicular to the interface (the $z$ direction) in parallel with our molecular dynamics simulations. This "dual simulation method" is particularly useful for regions where the potential of mean force is non-zero and one can no longer use the simple diffusion equation to solve for $D_{zz}$. The Langevin equation for motion along the $z$ direction is,

$$m\ddot{z} = -\frac{\partial W(z)}{\partial z} - \zeta_{zz}\dot{z} + R(t), \qquad (22)$$

where the first term on the right is the mean force on the particle (arising from the presence of the interface) derived from the potential of mean force $W(z)$, $R(t)$ is the random force which is assumed to be a Gaussian random process with a white noise spectrum, and $\zeta_{zz}$ is the static friction coefficient along the $z$-direction, which is related to the diffusion tensor through the Stokes-Einstein relation:

$$\zeta_{zz} = k_B T D_{zz}^{-1}. \qquad (23)$$

By comparing the survival probability, Eq. (15), determined from a full MD simulation with that determined from an analogous Langevin dynamics (LD) simulation, we can determine the best-fit value for the diffusion coefficient $D_{zz}$. The dual simulation method proceeds as follows. One computes the survival probability from both molecular dynamics and from a series of Langevin simulations corresponding to different values of the diffusion coefficient $D_{zz}$ for a region defined by $a < z < b$. The correct value of the diffusion coefficient is chosen to be the one that gives the best fit of the survival probability from LD to that from MD. As we show below, it will be necessary to make corrections to the positions of the virtual absorbing boundaries in the LD simulations.

Another strategy, bearing a close connection to the above but not used in this study, is to devise a dual simulation method in which one runs Smoluchowski dynamics (simulating the diffusion process defined by the Smoluchowski equation, unfortunately sometimes called Brownian dynamics) in parallel with full MD. In this case one would use virtual absorbing boundaries for the MD as defined above, but with relevant corrections to the boundaries for the Smoluchowski dynamics (SD) simulation as discussed below. In fact whenever one calculates the time dependence of dynamical properties that spring from the Smoluchowski or Langevin equation (LE), such as the corresponding autocorrelation function of its eigenfunctions or survival probability, it will be necessary to apply the relevant boundary corrections wherever the width of the layers appears explicitly.

### C. Two corrections to the positions of the virtual boundaries

It is important to recognize that many dual simulation methods will require corrections to the positions of the virtual absorbing boundaries.

*Smoluchowski dynamics versus Langevin dynamics*

In our molecular dynamics simulation of the air-water interface system the virtual absorbing boundary conditions are defined by the flux, $j_{out}(z = a, t) = 0$ and $j_{out}(z = b, t) = 0$, for a region defined within $a < z < b$. Virtual absorbing boundaries are also defined by the flux, $j_{out}(z = boundary, t) = 0$, in Langevin dynamics. However in the Smoluchowski equation absorbing boundaries are defined by the Smoluchowski boundary condition, $p(z = boundary, t|z_0, 0) = 0$. The discrepancy between the Smoluchowski boundary condition for absorption and the real absorbing boundary condition defined by the flux has been discussed at length.[13–15] Waldenstrom et. al.[13] have derived a correction factor for the width of the region, that equates the boundary condition $p(z = boundary, t|z_0, 0) = 0$ with the flux boundary condition $j_{out} = 0$:

$$l_{p=0} = l_{j_{out}=0} + 2\lambda, \tag{24}$$

$$\lambda = \left(\pi m D_{zz}^2 / 2 k_B T\right)^{1/2}, \tag{25}$$

and where $l_{p=0}$ is the corrected region width, necessary if one wants to use the Smoluchowski boundary condition to solve the diffusion equation, (or run SD), for a simulation with boundary condition $j_{out} = 0$ and width of length $l_{j_{out}=0} = L$.

*Langevin versus Generalized Langevin Dynamics*

Before proceeding we must determine if the static Langevin equation can accurately predict the decay of either the survival probability or the decay of the autocorrelation function of any eigenfunction of the diffusion operator (or Smoluchowski operator) for thin layers of liquid. There is an extensive literature showing that in real bulk liquids the generalized Langevin equation (GLE), not the static friction LE, describes the single particle dynamics. For example, the velocity correlation function exhibits marked deviations from the exponential decay predicted by the LE with static friction. This arises from the fact that the correlation time of the force experienced by a particle is of finite duration and not of zero duration as is implicit in the ordinary LE. The dynamic friction in the GLE has an important affect on the mean square displacement. The memory effect in the GLE leads to a mean-square displacement that initially increases faster than that predicted by the LE with white noise friction. However, in the long time limit, the rate of increase of the MSD is the same for both the GLE and LE with the same static limit of the friction coefficient.[16] The short time discrepancy becomes important when one considers properties dependent on the boundary such as the eigenfunction autocorrelation function or survival probability. Because the dynamics predicted by the GLE leads to faster translational motion at short time than would be predicted by the LE, particles described by the GLE initially distant from the boundaries eventually come close to the boundaries, cross and are absorbed earlier than they would be if their motion was described by the LE. In some sense we can regard the GLE as giving rise to a "jump distance", $\omega$, that would not be present in the LE simulation. Thus to a particle approaching an absorbing boundary the position of the boundary will appear to be closer by the length $\omega$ in GLE (or MD) as compared to LE. To correct for this in the dual simulation method we propose to use an effective layer width for the LE that is smaller than the width in the MD. This approach is similar in spirit to the correction factor derived by Waldenstrom cf Eq. (24), discussed above,

$$l_{LE} = l_{MD} - 2\omega, \tag{26}$$

where $l_{LE}$ is the corrected region width, (for two absorbing boundaries), necessary if one wants to use the static LE (or the diffusion equation) in the dual simulation method and the width of the region in MD is $l_{MD}$. We conjecture that this jump distance is equal to the maximum difference between $\sqrt{\langle \Delta z^2(t) \rangle}$ for the MD (or GLE) and LE simulations.

### III. THE APPLICATION TO THE VAPOR-WATER INTERFACE

#### A. The justification of the memory effect correction

To test the validity of our conjecture, we ran stochastic molecular dynamics based on the GLE with memory friction $\zeta_{zz}(t)$:[17]

$$m\ddot{z} = -\frac{\partial W(z)}{\partial z} - \int_0^t d\tau \zeta_{zz}(t-\tau)\dot{z}(\tau) + R(t). \tag{27}$$

The Generalized Langevin dynamics (GLD) simulations were conducted with an exponential memory kernel $\zeta_{zz}(t) = \zeta_0 \alpha e^{-\alpha t}$, where $\zeta_0$ is the static friction kernel corresponding to Eq. (23), and $1/\alpha$ is the memory relaxation time. For a system with fictitious absorbing boundaries at positions $a$ and $b$ where $b - a = 3.4 \text{Å}$, the survival probability is calculated. Langevin dynamics with the same static friction kernel and absorbing boundaries at positions $a + \omega$ and $b - \omega$ were run varying $\omega$ until the decay of the survival probability coincide for LD and GLD. The maximum difference in $\sqrt{\langle \Delta z(t)^2 \rangle}$ for GLD and LD is computed and plotted in the Fig.[ 2](a) along with $\omega$ as a function of the memory relaxation time. The two curves fit with each other quite well, an observation that supports our conjecture concerning the adjusted boundaries.

To provide further evidence for the conjecture, we run a $400ps$ molecular dynamics simulation of a homogeneous isotropic sample of liquid water consisting of 256 TIP4P[18] water molecules with periodic boundary conditions. From the velocity autocorrelation function, we adopt the method of Berne and Harp[19] to calculate the friction kernel numerically, as shown in Fig. [ 2](b). We determined the maximum difference in $\sqrt{\langle \Delta z(t)^2 \rangle}$ between GLD( with this empirical friction kernel) and LD (with the corresponding static friction coefficient), giving a value of $\omega = 0.16$ Å. A plot of the survival probability, $P(t)$, for this GLD with the empirical kernel, for the original MD and for LD with width corrected by $\omega = 0.16$ Å is given in Fig.[ 2](e). The close correspondence between $P(t)$, determined from LD with the corrected width $l_{LE}$, and $P(t)$ determined from either MD or GLD ( with a friction kernel found from the MD) further justifies are conjecture. [Agreement between the original MD and GLD velocity autocorrelation functions and mean-square displacements, and between the random force autocorrelation function used in the GLD and the friction kernel found from MD shown in Fig. [ 2](b)(c)(d), demonstrates the accuracy of this test of our above conjecture.] Given the validity of this conjecture, we can now fit the friction coefficient in LD simulations with corrected boundary conditions to MD using the dual method described before. We do not need to determine the explicit time dependence of the friction kernel.

### B. The parallel components

Before applying Eq. (16) to a liquid with an interface, we first test this approach on the same bulk water system used in our justification of the correction factor. We compute the diffusion coefficients in slabs of water of width $3.5 \text{Å}$ perpendicular to the z-axis using Eq. (16). This method gives $D_{xx} = 0.30 \pm 0.02$ Å$^2$/ps, $D_{yy} = 0.30 \pm 0.02$ Å$^2$/ps. We also used the Einstein relation (c.f. Eq. (1)), in the standard way, to compute the diffusion coefficient for this homogeneous system and found it to be $D = 0.307$Å$^2$/ps in agreement with our calculation, therefore validating this methodology.

Liquid water with a liquid-vapor interface is prepared using 512 water molecules of the TIP4P/FQ[18,3] model. The resultant 3-D periodic simulation box is rectangular with dimensions $L_x = L_y = 25$ and $L_z = 75$. The width of the water phase is approximately $25 \text{Å}$ and is centered about the middle of the box $z = 0$. Further simulation details are given in the Fig.[ 1] caption. Two water-vapor interface regions perpendicular to the $z$ coordinate are present on either side of $z = 0$. The liquid is subdivided into layers of water of width $\Delta z = 3.5 \text{Å}$ perpendicular to the $z$-axis. The interface width is $3.4 \text{Å}$ and the center of the region is positioned at $z = 12.46 \text{Å}$. We evaluate $D_{xx}$ and $D_{yy}$ in each of the layers of the liquid from Eq. (16). Plots of $\langle \Delta x^2(t) \rangle / 2p(t)$ and $\langle \Delta y^2(t) \rangle / 2p(t)$ for the slab in the middle of the bulk water phase and in the interface layer are given in Fig.[1]. This figure also show the range over which the linear fitting is applied, from $2.0 - 3.0 ps$. The plot of $D_{xx}$, $D_{yy}$ ranging from the middle of the bulk phase to the interface is given in Fig.[ 1]. The error bars are calculated using block averaging over the trajectory.[2]

### C. The perpendicular component

For finite regions that border on the interface with water vapor, the potential of mean force is significantly different from zero and the long time diffusional motion will be given by the Smoluchowski equation, Eq. (6). For the purposes of the following discussion we define the liquid side boundary at $z = l$ and the vapor side boundary at $z = 0$. To use the methods outlined before, we assign fictitious boundaries that are perfectly absorbing at $z = 0, l$. Fig 1 shows how the density varies as a function of position as one approaches the interface from inside the liquid as determined from MD. This profile can be fit to the hyperbolic tangent function,[20] as

$$\rho(z) = \frac{1}{2}\left[1 + \tanh[2.1972(z - l/2)/l]\right]\rho_0 \tag{28}$$

for the interfacial region. The potential of mean force in our water vapor interface is then defined as

$$W(z) = -kT \ln\left(\rho(z)/\rho_0\right). \tag{29}$$

Needless to say, it is difficult to solve the Smoluchowski equation with this non-linear potential of mean force analytically, thus necessitating the use of the dual LD simulation method for this region.

We have outlined two routes to determining the $D_{zz}$ component of the diffusion tensor. The first involves using the analytical expression, given by Eq. (20), for the decay of the autocorrelation of the eigenfunction of the Smoluchowski Equation. When the solution of the Smoluchowski Equation is difficult to obtain, a second route, using dual simulation Langevin Dynamic simulations can be used to evaluate $D_{zz}$.

Before applying the analytical solution, Eq. (20) and the LD dual simulation method to a liquid with an interface, we first test this approach on the same homogeneous, isotropic sample of liquid water used in the calculation of $D_{xx}$ and $D_{yy}$. We compute the diffusion coefficients in slabs of water of width $3.5 \mathring{A}$ perpendicular to the z-axis using Eq. (20). The autocorrelation of the eigenfunction or the survival probability is calculated from the MD simulation for a region defined by $a < z < b$ where $L = b - a = 3.5 Å$. Using the diffusion equation and thus Eq. (20) to solve for $D_{zz}$ requires two corrections to the width $L$. As discussed above the discrepancy between the Smoluchowski boundary condition ($p = 0$) used to solve the diffusion equation and the real boundary condition ($j_{out} = 0$) can be corrected using $\lambda$ defined in Eq. (24). Secondly one must include the aforementioned memory correction $\omega$ giving an effective width $l_{eff} = 3.5 + 2\lambda - 2\omega$, where $\lambda = 0.104 Å$ and $\omega = 0.143 Å$ for this system. This method gives $D_{zz} = 0.30 \pm 0.02$ $Å^2/ps = D_{xx} = D_{yy}$. Using the Langevin dual simulation method requires only the memory correction leading to absorbing boundaries in the LD simulation at $a + \omega$ and $b - \omega$. From the dual Langevin method $D_{zz} = 0.30 \pm 0.02$ $Å^2/ps$ in perfect agreement. The diffusion tensor is found to be isotropic, as we expect for this system with values in agreement with the correct diffusion coefficient calculated using the Einstein relation (c.f. Eq. (1)), in the standard way, $D = 0.307 Å^2/ps$. Thus our new method based on analyzing the MD data by imposing virtual absorbing boundary conditions gives the same value for the diffusion coefficient as the standard method for the homogeneous system, thus vindicating the method.

To evaluate $D_{zz}$ for the bulk water like regions ($z < 10 Å$) of our air-water interface system we compute the autocorrelation of the eigenfunction with eigenvalue $n = 1$ and fit to the exponential in Eq. (20), (the results from dual simulation are equivalent for the bulk like regions). A plot of the fit for a $3.5 Å$ region centered in the middle of the bulk water phase is given in Fig.[ 3](a).

We now apply the above LD dual simulation analysis to determine the components of the diffusion tensor in the interfacial layers of our air-water interface system. The dual simulation method is a little more complicated than the procedure for the previously discussed bulk water like regions. The correction factor ($\omega$) is derived from the maximum difference of $\sqrt{\langle \Delta z(t)^2 \rangle}$ between MD and LD for the same value of the diffusion coefficient. Not knowing the value of the diffusion coefficient on the interface we resort to an iterative procedure. Initially we use dual simulation to solve for $D_{zz}$ without considering $\omega$. We then calculate $\omega$ from the corresponding MSD curves from MD and LD. Using this value of $\omega$ we change the position of the boundaries in the LD simulation and derive a new $D_{zz}$ using dual simulation. This procedure is repeated until the change in the diffusion coefficient is smaller than $0.01 Å^2/ps$. A plot of the final fit for the interface region is given in Fig.[ 3](b). The plot of $D_{zz}$ and the previously calculated $D_{xx}$ and $D_{yy}$ ranging from the middle of the bulk phase to the interface is given in Fig.[ 1]. The error bars are calculated using block averaging over the trajectory.[2]

Implicit in the MD simulation is the $z-$dependent interfacial potential of mean force. Because of this the ordinary Einstein relation and the Green-Kubo equation for the diffusion coefficient (the time integral of the velocity autocorrelation function (VAF)) are no longer valid. Consider, for example, Langevin dynamics along $z$ with the systematic force arising from the interfacial potential of mean force and with a given static friction coefficient $\zeta_{zz}$. In general the VAF will be functionally dependent on the PMF as will be its integral, i.e. the presumed diffusion coefficient, will also be functionally dependent on the PMF. In contrast, for the Smoluchowski equation, the diffusion coefficient is independent of the PMF. Disregarding this, one might wonder what the value of $D_{zz}$ would be if applying the Einstein equation on the interface was correct. Using Eq. (1) for particles initially in the same interface region used above, we get $D_{zz} = 0.23 \pm 0.02$ Å$^2$/ps. The value for $D_{zz}$ is clearly much smaller than found from the dual simulation method. Integrating the first 3 ps of the Green-Kubo relation gives $D_{xx} = 0.8 \pm 0.5$ Å$^2$/ps, $D_{yy} = 0.7 \pm 0.5$ Å$^2$/ps and $D_{zz} = 0.3 \pm 0.5$ Å$^2$/ps. One may argue that the uncertainties encountered are too large to make a definitive statement. To clarify the inadequacy of using the Green-Kubo relation to compute $D_{zz}$, we performed a very long LD simulation, like the one used in the dual simulation method of the interfacial region. This mimics the MD. We determine the $z$-component of the VAF for a static friction consistent with $D_{zz} = 0.52$ Å$^2$/ps. Integrating the first 10 ps of the VAF gives an erroneous value of $D_{zz} = 0.23 \pm 0.05$ Å$^2$/ps, approximately half as large as the input value.

## IV. DISCUSSION AND CONCLUSION

Far from the interface the diffusion tensor is found to be isotropic, as expected, and the diffusion coefficient has the value $D \approx .22 Å^2/psec$ in agreement with what is found in the bulk liquid.[3] As the layers approach the interface region all components of the diffusion tensor in-crease. This can be understood when one considers the source of water's relatively slow mobility in bulk water. In bulk water each molecule forms on average 3.6 hydrogen bonds with neighboring water molecules. This hydrogen bonding network impedes the translational motion of the water molecules. In the interface region the number of hydrogen bonds is approximately 2.5 per water molecule.[10] The fewer hydrogen bonds reduce the effective friction felt by the water molecules resulting in the larger diffusion coefficient. The $D_{zz}$ component of the diffusion coefficient of the interface is approximately two times the value in bulk water, ($\approx .5 \text{Å}^2$/psec) while the components parallel to the interface ($D_{xx} = D_{yy} \approx .8 \text{Å}^2$/psec) are approximately three and a half times the bulk value. The axial anisotropy in the diffusion is related to the structural asymmetry of the interface.

We suspect that the same qualitative behavior will be seen in all liquids, including classical liquids such as liquid argon. In fact our simulations of liquid argon at reduced temperature and density $0.75$ and $0.83$ gave $D_{xx} = D_{yy} = D_{zz} \approx 0.2 \ \text{Å}^2/psec$ in the bulk and $D_{xx} = D_{yy} \approx 0.8 \ \text{Å}^2/psec$ and $D_{zz} \approx 0.4 \ \text{Å}^2/psec$ in the interfacial layer.

One can rationalize this behavior as follows. The barriers for diffusion perpendicular to the interface will be larger than for diffusion parallel to the barrier, but will be smaller than for diffusion in the bulk. A vacancy model for diffusion illustrates this. For diffusion to occur, a vacancy must be next to the diffusing particle. Since fluid relaxes around this vacancy, the barrier opposing a jump into the vacancy depends on the "stiffness" of the fluid with respect to density fluctuations transverse to the direction of the jumping particle. For a jump into a vacancy along the x of y direction, we must consider the stiffness along the z direction, whereas for a jump into the vacancy along the z direction, we must consider the stiffness along the x or y directions. In the interface we expect the stiffness along z to be smaller than along x and y. Thus the barrier to diffusion along z will be greater than the barrier to diffusion along either x or y, and $D_{zz} < D_{xx} = D_{yy}$. Likewise we expect that all barriers in the bulk will be larger than in the interface because the fluid will be stiffer there. This hand-waving argument is consistent with our observations.

In summary, we have developed a general method to calculate the self-diffusion coefficient in a finite region. This is very important since the mobility of the solvent molecules greatly effect the reactivity and dynamics of solutes such as proteins. In principle, our method can be extended to more complicated geometries and different boundary conditions, to handle problems that include the diffusivity of water in the ion channel or 'pocket' near the active site in some enzymes.

## ACKNOWLEDGMENT

This paper is dedicated by B.J.B. to Hans C. Andersen on the occasion of his sixtieth birthday. Hans has been a valued colleague and friend for more than thirty years. This work was supported by a grant to B.J. Berne from NSF (CHE-03-16896).

FIGURES

FIG. 1. (a) In order to calculate the components of diffusion parallel to our boundary surface we use the generalized Einstein equation for the anisotropic diffusion equation (16). The bulk region is defined as a $3.5 Å$ region in the center of our water phase,The interface region is defined in the text with a width of $3.4 Å$. The portions of the curve we fitted is between $2.0 ps$ and $3.0 ps$. (b) Plot of the diffusion coefficient in all three Cartesian coordinates for $3.5 Å$ regions extending from the bulk up to the interface. A $400 ps$ trajectory is run in the NVE ensemble using the velocity-verlet integrator to update the atomic positions and velocities. The atomic configuration is recorded for data analysis every $20 fs$.The coordinate correction algorithm, RATTLE,[21] is used to constrain the bond lengths allowing for a timestep of 1 fs in our simulations. To efficiently calculate the electrostatic interactions, mesh based approximations to the Ewald sum[22] can be used.[23–26] We use the P3ME[23] method.

FIG. 2. (a)The plot of memory correction factor vs. $\alpha$ (inverse of characteristic time). The dash line and solid line is the correction factor $\omega$ from the GLD and LD and the biggest difference of $\sqrt{\langle \Delta z(t)^2 \rangle}$, respectively. The static friction corresponds to a $D = 0.25 Å^2/ps$, $T = 300 K$ and water molecule mass. (b) The friction kernel. The solid line is calculated from the velocity autocorrelation function of MD by solving the Volterra equation numerically, see cited reference[19] for details. From the second fluctuation-dissipation theorem, the relationship between the friction kernel and the random force autocorrelation function (FAC) can be established as $\zeta_{zz}(t) = \beta \langle R(0)_z R(t)_z \rangle$, where $\beta = 1/(kT)$ and $R_z$ is the random force. The curve of FAC is scaled by $\beta$. (c) The velocity autocorrelation function (VAC) from the molecular simulation and the GLD. (d) The mean-square displacement. The GLD simulation with real friction kernel can almost reproduce the same MSD curve as the real MD simulation. The difference of the $\sqrt{\langle \Delta z(t)^2 \rangle}$ is obvious.(e) The survival probability from MD, GLD with real friction kernel and Langevin dynamics with width corrected using mean square displacement prescription. The coincidence of these curves proves the accuracy of the numerical kernel and the memory correction.

FIG. 3. (a)The decay of the autocorrelation function of the eigenfunction for the first non-zero eigenvalue is fit using a least squares method for the bulk region. See Eq. (20). (b) The decay of the survival probability for the interface region. The survival probability is fit for long times using a dual simulation with Langevin Dynamics.

Mean Square Displacement for Interface and Bulk

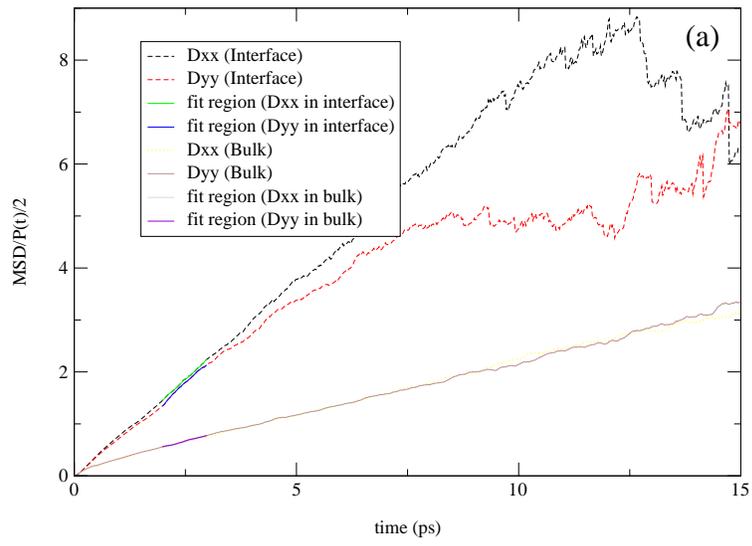

z-dependance of Diffusion Coefficient

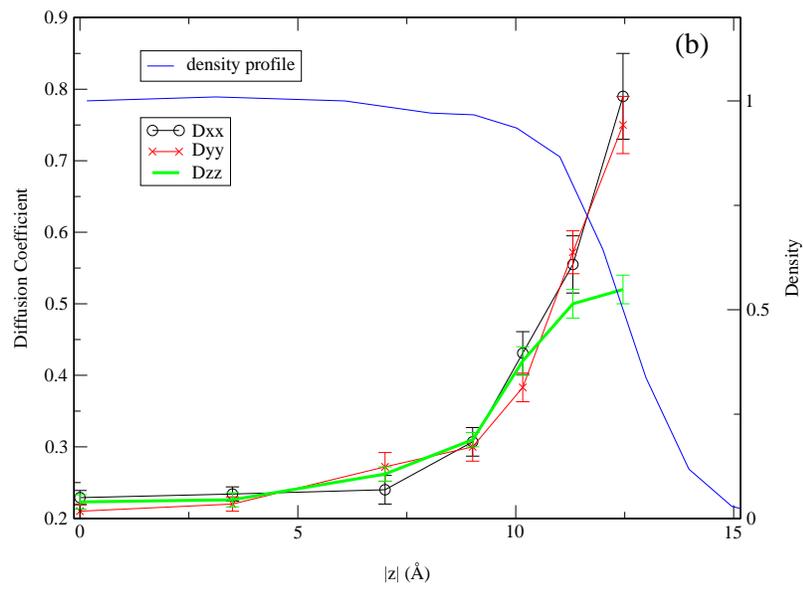

Fig 1.

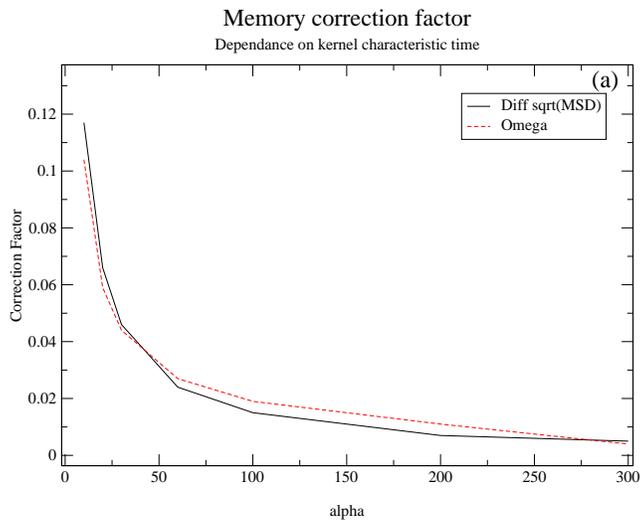
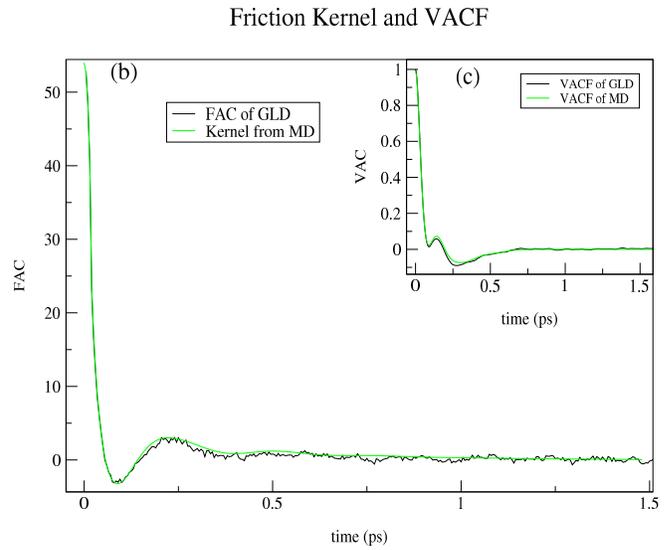
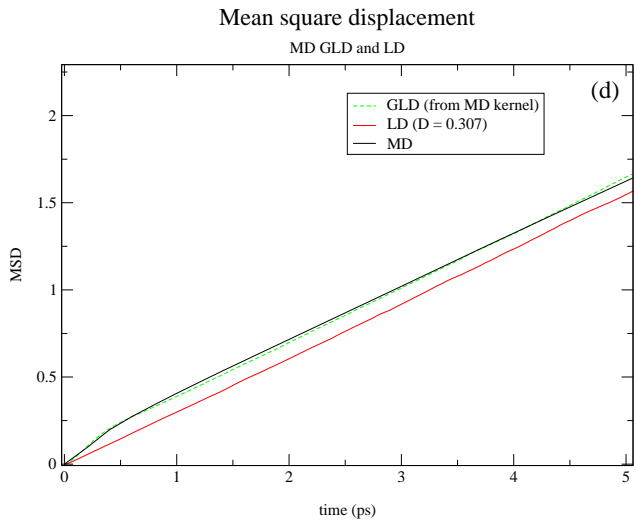
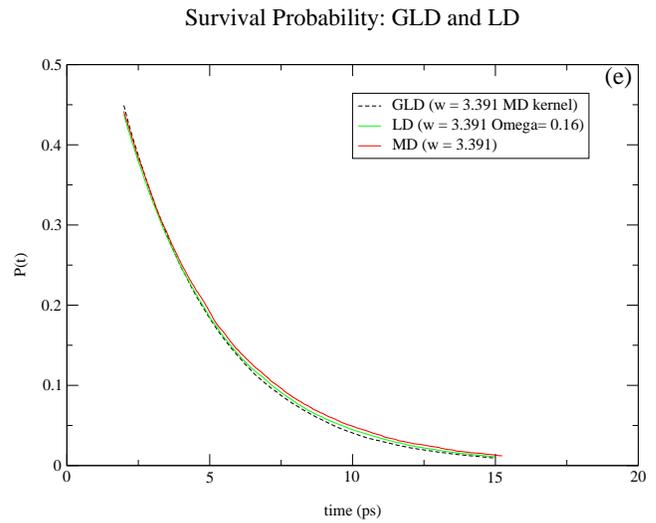

Fig 2.

## Eigenfunction autocorrelation function & Survival Probability

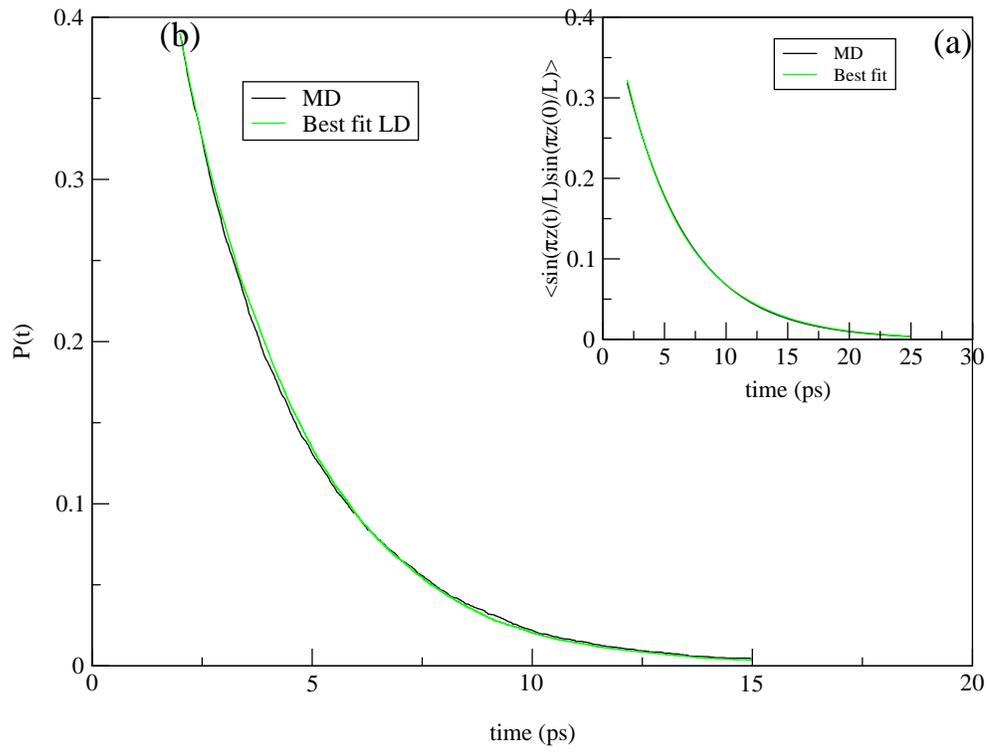

Fig 3.